\documentclass[twocolumn,showpacs,oneside,prb]{revtex4}
\usepackage{epsfig,amssymb,amsfonts}

\begin{document}

\title{Spatial Reflection and Associated String Order in Quantum Spin Chains}
\author{Li-Xiang Cen}
\affiliation{Department of Physics, Sichuan University, Chengdu
610065, China}

\begin{abstract}
We investigate spatial reflection and associated nonlocal order in
spin chain quantum systems. The proposed string order parameters,
e.g., reflected via operations of the spatial reflection or
combinations of it with spin reflection, are able to characterize a
variety of physical systems and allow us to gain renewed insights to
the statistical mechanism underlying phenomena such as the Haldane gap
and quantum phase transitions. Besides revealing further the
potential application of the generalized parity symmetry in
numerical algorithm, we build an explicit scheme to determine the
symmetry and the related string order for matrix product states so
that one can construct ansatz models with presumed properties.
\end{abstract}


\maketitle

Order parameter has been one of the most important concepts in
condensed matter physics since Landau's finding of it in describing
continuous phase transitions with spontaneous symmetry breaking. The
nonlocal string order, originally introduced by den Nijs and
Rommelse \cite{den}, was thought of a hidden antiferromagnetic
N\'{e}el order and exploited to understand the mechanism of the
Haldane gap \cite{haldane}. Despite that Haldane ground states of
integer-$S$ Heisenberg chains possess only short-range spin-spin
correlations, they could have nonzero string order parameters. The
latter was found to be a common feature of the Haldane phase
resulted from breaking of certain topological symmetry \cite{symme}
and the phenomenon has been investigated for various physical models
\cite {hida}. A typical example is the valence bond solid (VBS)
model \cite{AKLT} and its higher-$S$ generalizations \cite{sun,son},
in which the string order parameter can be explicitly worked out.

To determine the hidden order for a general quantum many-body system
is a problem highly nontrivial. In a recent literature
\cite{string}, the presence of den Nijs-Rommelse string order is
demonstrated rigorously in relation to the existence of local
symmetries within the framework of matrix product states (MPSs). The
revealed connection is somewhat universal in the sense that the MPS
formulation indeed offers a general mathematical representation for
quantum many-body states \cite{fannes,vidal,faith}. As a
consequence, the problem of determining the den Nijs-Rommelse string
order for a quantum system is recast as that of finding out possible
local symmetries in the system. At this stage, it is of interest to
build a scheme to identify the local symmetry for general MPSs,
since for every MPS with a finite representative dimension an ansatz
model with local interactions can always be constructed
\cite{fannes} such that the given MPS constitutes its ground state.

In this paper we investigate the spatial reflection and propose
novel string order to characterize spin lattice systems. Differing
from conventional string operators utilizing local unitary
transformations, the present string operator acting on spins in
between two boundaries is defined by the spatial reflection or
combinations of it with local spin reflection. By virtue of the MPS
representation, the proposed string order is shown to relate
intimately to the symmetry with respect to parity or generalized
parity transformations. The quantity is then applied to characterize
a sort of models including the VBS states and a spin-$\frac 12$
system with quantum phase transitions. Similar to the parity
symmetry, the generalized parity symmetry could also be applied to
reduce computational costs in numerical algorithm of the density
matrix renormalization group (DMRG) \cite{white,umod}. In addition,
we build an explicit protocol to determine from first principles the
string order for MPSs with representative dimension $D=2$ and
present an example of modeling ansatz with presumed properties.

The primitive form of the string order parameter employed here is
proposed as
\begin{equation}
{\rm SO}_P^\alpha \equiv \lim_{|j-i|\rightarrow \infty }\langle
S_i^\alpha \mathcal{P}^{(l)}S_j^\alpha \rangle ,  \label{sop}
\end{equation}
where $S_i^\alpha $ ($\alpha =x,y,z$) denotes the spin operator of
the $i$th lattice site and the parity operator acts on $l=j-i-1$
consecutive sites between the boundary spins $i$ and $j$ as
$\mathcal{P}^{(l)}|s_{i+1}\cdots s_{i+l}\rangle =|s_{i+l}\cdots
s_{i+1}\rangle $. This string operator and its generalized form of
Eq. (\ref{SOGP}) can be exploited to characterize the hidden order
for infinite lattice systems just as the original den Nijs-Rommelse
string order.

To be specific, let us resort to the MPS formulation for
translationally invariant spin chains
\begin{equation}
|\Psi \rangle =\frac 1{\sqrt{TrE^N}}\sum_{\{s_i\}}Tr(A^{s_1}\cdots
A^{s_N})|s_1,\cdots ,s_N\rangle ,  \label{form}
\end{equation}
where $s_i=1,\cdots ,d$ specifies the lattice spin degrees of
freedom and $\{A^s\}$ is a set of $D\times D$ matrices
parameterizing the state $|\Psi \rangle $. The transfer matrix $E$
contained in the normalization factor is given as
$E=\sum_{s=1}^d(A^s)^{*}\otimes A^s$. By using further the
notations, say, $E_{S^\alpha }\equiv \sum_{s,s^{\prime }}\langle
s|S^\alpha |s^{\prime }\rangle (A^s)^{*}\otimes A^{s^{\prime }}$,
the quantity ${\rm SO}_P$ of the state $|\Psi \rangle $ is shown to
be
\begin{equation}
{\rm SO}_P^\alpha =\lim_{l\rightarrow \infty }\lim_{N\rightarrow
\infty }\frac{Tr[E_{S_i^\alpha }E_{\mathcal{P}^{(l)}}E_{S_j^\alpha
}E^{N-l-2}]}{TrE^N}, \label{sopp}
\end{equation}
where
\begin{equation}
E_{\mathcal{P}^{(l)}}=[(E^{T_2})^l]^{T_2},~~E^{T_2}=\sum_s(A^s)^{*}\otimes
(A^s)^T.  \label{transfer2}
\end{equation}
Note that the state $|\Psi \rangle $ is identified to be parity
symmetric \cite{ostlund,cen} iff there is $\langle \Psi
|\mathcal{P}^{(N)}|\Psi \rangle =Tr(E^{T_2})^N/TrE^N=\pm 1$. The
latter means that for an infinite spin chain the largest eigenvalue
$\lambda _m^{\prime }$ of $E^{T_2}$ should have the same modulus
with the $\lambda _m$ of $E$. This fact together with the relation
$\lim_{l\rightarrow \infty }E_{\mathcal{P}^{(l)}}=\lambda _m^{\prime
l}(|\Phi _m^R\rangle \langle \Phi _m^L|)^{T_2}$ lead to that the
quantity ${\rm SO}_P^{\alpha }$ could be nonzero iff the state
$|\Psi \rangle $ is parity symmetric and the corresponding value is
calculated as
\begin{equation}
{\rm SO}_P^\alpha =\frac 1{\lambda _m^2}\langle \Psi
_m^L|E_{S_i^\alpha }(|\Phi _m^R\rangle \langle \Phi
_m^L|)^{T_2}E_{S_j^\alpha }|\Psi _m^R\rangle ,
\end{equation}
where $|\Psi _m^{L(R)}\rangle $ and $|\Phi _m^{L(R)}\rangle $ denote
respectively the left (right) eigenvectors of $E$ and $E^{T_2}$
related to $\lambda _m$.

Let us address a concrete example of the VBS ground state of the
Affleck-Kennedy-Lieb-Tasaki model \cite{AKLT} represented by
$\{A^s\}=\{\sigma _z,\sqrt{2}\sigma _{+},\sqrt{2}\sigma _{-}\}$. In
this case, $E$ and $E^{T_2}$ are Hermitian thus their dual bases of
left and right eigenvectors recover simply the conventional
orthonormal bases. Their largest eigenvalue $\lambda _m=3$ and the
corresponding eigenvectors are $|\Psi _m\rangle =\frac 1{\sqrt{2}
}(|00\rangle +|11\rangle ) $ and $|\Phi _m\rangle =$ $\frac
1{\sqrt{2}}(|01\rangle -|10\rangle )$. The string order ${\rm SO}_P$
is worked out to be $|{\rm SO}_P^\alpha |=\frac 1{18}$, which is
independent of the orientation $\alpha =x,y$ or $z$ owing to the
SU(2) symmetry of the model. Moreover, the extended model of the
SO(5) matrix product state \cite {son,zhang} specified by
\begin{equation}
A^0=\sigma _z\otimes \sigma _z,~A^{\pm 1}=\sqrt{2}\sigma _z\otimes
\sigma _{\pm },~A^{\pm 2}=\sqrt{2}\sigma _{\pm }\otimes I
\label{so5}
\end{equation}
has also nonzero ${\rm SO}_P$. The corresponding matrices $E$ and
$E^{T_2}$ are Hermitian again and have the largest eigenvalues
$\lambda _m=5$ with eigenvectors as
\begin{equation}
|\Psi _m\rangle =\frac 12\sum_{i=1}^4|ii \rangle ,|\Phi _m\rangle =\frac
12(|14\rangle -|41\rangle +|32\rangle -|23\rangle ).  \label{so5eigen}
\end{equation}
In terms of the $S^z$ quantum number of the $s=2$ lattice spins, the
string order is obtained as ${\rm SO}_P^z=-\frac 15$.

For further application let us consider a spin-$\frac 12$ matrix
product system specified by $A^0=$ (\negthinspace {\tiny $
\begin{array}{ll}
0 & 0 \\
1 & 1
\end{array}
\!\!$}), $A^1=$(\negthinspace {\tiny $
\begin{array}{ll}
1 & g \\
0 & 0
\end{array}
\!\!$}). This is an ansatz model \cite{wolf} described by a parent
Hamiltonian with three-body interactions and it undergoes a quantum
phase transition at the point $g=0$. The system is parity symmetric
and the matrices $E$ and $E^{T_2}$ have the same eigenvalues
$\lambda _{\pm}=1\pm g$. By definition, the string order parameter
${\rm SO}_P^z$ of the ground MPS is shown to take a discrete form
with ${\rm SO}_P^z=\frac 18$ as $g<0$ and ${\rm SO}_P^z=\frac 18
\left( \frac {1-g}{1+g}\right)^2$ as $g>0$. On the other hand, the
system possesses also a local $Z_2$ symmetry with respect to the
spin flip $|0\rangle \leftrightarrow |1\rangle $. The transverse den
Nijs-Rommelse string order associated with the local unitary
$U=\exp{(i\pi S^x)}$, namely,
\begin{equation}
{\rm SO}_D^{x}\equiv \lim _{|j-i|\rightarrow \infty}\langle S_i^{x}
U^{\otimes l}S_j^{x}\rangle ,~~l=j-i-1, \label{sod}
\end{equation}
is obtained as $|{\rm SO}_D^x|=g/(1+g)^2$ or ${\rm SO}_D^x=0$ for
cases $g>0$ and $g<0$, respectively. Thus the quantum phase
transition of the system crossing the point $g=0$ is marked
distinctly by the non-analytical behavior of both these string order
parameters.

As the quantity ${\rm SO}_P$ reveals the hidden order for states
with parity symmetry, we show below that the state with generalized
symmetry with respect to combination of the parity and spin
reflection indicates another type of nonlocal order. For a simple
example consider an MPS \cite{cen} represented by $\{A^s\}=$
\{(\negthinspace {\tiny $
\begin{array}{ll}
1 & 0 \\
0 & \sqrt{2}
\end{array}
\!\!$})$,$(\negthinspace {\tiny $
\begin{array}{ll}
0 & 0 \\
\sqrt{2} & 0
\end{array}
\!\!$})$,$(\negthinspace {\tiny $
\begin{array}{ll}
0 & 1 \\
0 & 0
\end{array}
\!\!$})\}. The state is parity absent but invariant under the
combined operation $\mathcal{P}^{(N)}U_P^{\otimes N}$, where $U_P$
is a spin flip operator specified as $U_P=|1\rangle \langle
1|+|2\rangle \langle 3|+|3\rangle \langle 2|$. This sort of
generalized parity symmetry is identified for MPSs as there is
\begin{equation}
(A^i)^T=\sum_{j=1}^dU_P^{ij}(XA^jX^{-1}),~~i=1,\cdots ,d
\label{localeq}
\end{equation}
and it is readily verified that for the above example there exists
$X=~$(\negthinspace {\tiny $
\begin{array}{ll}
\sqrt{2} & 0 \\
0 & 1
\end{array}
\!\!$}). Indeed, the kind of symmetry was already known in fermion
systems in performance of the DMRG procedure \cite {umod}. Owing to
the fermion anti-commutation rule, the Jordan-Wigner transformation
will result in a parity asymmetric Hamiltonian which is, however,
invariant under the combination of the parity and a simple spin
reflection.

Accordingly, one is led to consider a string operator
\begin{equation}
{\rm SO}_{GP}^\alpha \equiv \lim_{|j-i|\rightarrow \infty }\langle
S_i^\alpha \mathcal{P}^{(l)}U_P^{\otimes l}S_j^\alpha \rangle ,
\label{SOGP}
\end{equation}
where $\mathcal{P}^{(l)}U_P^{\otimes l}$ describes the combined
transformation on $l=j-i-1$ consecutive lattice sites in between $i$
and $j$. For an MPS with generalized parity symmetry specified by
Eq. (\ref{localeq}), this string quantity is shown to be
\begin{equation}
{\rm SO}_{GP}^\alpha =\frac{\langle \Psi _m^L|E_{S_i^\alpha
}V^{-1}(|\Psi _m^R\rangle \langle \Psi _m^L|)^{T_2}VE_{S_j^\alpha
}|\Psi _m^R\rangle }{\lambda _m^2},  \label{sogpp}
\end{equation}
where $V\equiv I\otimes X$, $|\Psi _m^{L,R}\rangle $ are
eigenvectors of $E$ related to $\lambda _m$, and the relation
$\lim_{l\rightarrow \infty }(E^l)^{T_2}=\lambda _m^l(|\Psi
_m^R\rangle \langle \Psi _m^L|)^{T_2}$ has been applied to obtain
the equality. The string order of the MPS mentioned above Eq.
(\ref{localeq}) is then derived as ${\rm SO}_{GP}^z=\frac 7{81}$ and
${\rm SO}_{GP}^x=-{\rm SO}_{GP}^y=\frac 4{81}$.

Revelation of the existence of the generalized parity symmetry in
spin lattice systems is important in many aspects. For instance, it
can be exploited to reduce computational costs in numerical
algorithm. In fact, it has long been realized in the DMRG algorithm
\cite{white} that for systems with parity symmetry the density
matrix of environment blocks could be achieved via a simple
reflection on that of the system blocks. The various existence of
generalized parity symmetries can also be utilized to achieve a
factor $\sim 2$ speedup and the density matrices of system and
environment blocks are now connected via the generalized parity
transformation. Moreover, in view that an MPS can be viewed as a
ground state of an ansatz system with short-range interactions, it
is of interest to construct the parent Hamiltonian of the modeling
system with certain presumed parity symmetry. The latter implies
that the derived system might possess nonlocal order associated with
the specified string operator.

We now focus on the problem of determining the generalized parity
symmetry for MPSs so as to construct ansatz lattice models with
prescribed properties. The question arises as: given an MPS $|\Psi
\rangle $, how to certify the relation (\ref {localeq}) exists or
not and how to find out such $U_P $ and $X$ if they do exist? To
this end, we employ an expression equivalent with Eq. (\ref{localeq}) but
in terms of the transfer matrix
\begin{equation}
E^T=(X^{*}\otimes X)E(X^{*}\otimes X)^{-1}.  \label{localequ}
\end{equation}
In comparison with Eq. (\ref{localeq}), this equation focuses solely
on the invertible matrix $X$ and the task reduces to distinguish a
particular $S$, among all those satisfying $E^T=SES^{-1}$, that
could be decomposed into $S=X^{*}\otimes X$. The difficulty of the
problem comes from that the similar transformation $S$ connecting
$E^T\sim E$ is not unique and the complexity even increases with
representative dimensions. Intriguingly, we show below that for
cases of $D=2$ the problem could be resolved by virtue of a
particular realization of group homomorphism between the direct
product group of unimodular linear transformations and the
four-dimensional complex orthogonal group \cite{cen1}.

Briefly, it was shown in Ref. \onlinecite{cen1} that in terms of a
sort of so-called pseudo-orthonormal bases, the representative
matrix of the operator $X_1\otimes X_2$, where $X_{1,2}$ denote
arbitrary linear transformation of SL(2,C), is complex orthogonal
and constitutes a group element of SO(4,C). The result leads to an
intuitive perception that the pseudo-orthonormal bases could be
utilized to resolve the factorization problem for the mentioned
similar transformation $S$. To be specific, the following simplest
pseudo-orthonormal bases (the so-called ``magic bases'', see Ref.
\onlinecite{bennett}) will be employed
\begin{eqnarray}
|e_1\rangle &=&\frac{\sqrt{2}i}2(|00\rangle +|11\rangle ),|e_2\rangle =
\frac{\sqrt{2}}2(|00\rangle -|11\rangle ),  \nonumber \\
|e_3\rangle &=&\frac{\sqrt{2}}2(|01\rangle +|10\rangle ),|e_4\rangle
=\frac{\sqrt{2}i}2(|01\rangle -|10\rangle ).  \label{magic}
\end{eqnarray}
Suppose that the transfer matrix has full rank and is expressed as
$E=\sum_{i=1}^4\lambda _i|\Psi _i^R\rangle \langle \Psi _i^L|$
according to its spectrum structure. Its transposition is then
formed as $E^T=\sum_{i=1}^4\lambda _i|\Psi _i^{L*}\rangle \langle
\Psi _i^{R*}|$ and the series of similar transformations connecting
them are given by $S(k_i)=\sum_{i=1}^4k_i|\Psi _i^{L*}\rangle
\langle \Psi _i^L|$, where the parameters $k_i\neq 0$ are to be
determined later. The distinguishing problem of Eq. (\ref{localequ})
could be resolved according to the following protocol: i) Express
the operator $S(k_i)$ in terms of bases (\ref{magic}) and denote the
representative matrix as $D_{\mu \nu }^s(k_i)=\langle e_\mu
|S(k_i)|e_\nu \rangle $; ii) Presume that $D^s(D^s)^T=(D^s)^TD^s=I$
and resolve the set of $k_i$ if they do exist; iii) Factorize the
operator as $S(k_i)=X_1\otimes X_2$ according to the correspondence
of the group homomorphism ${\rm SO(4,C)}\sim {\rm SL(2,C)}\otimes
{\rm SL(2,C)}$ and verify if there exists $X_2=X_1^{*}$. Once $X$ is
figured out, the spin reflection $U_P$ can be worked out easily from
Eq. (\ref{localeq}).

It is worthy to note that a slightly modified scheme is capable to
determine the local symmetry hence the string order $SO_D$ for MPSs,
wherein the condition reads as \cite {string}
$A^i=\sum_jU^{ij}(XA^jX^{-1})$. The problem then becomes to find out
a decomposable transformation $S=X^*\otimes X$ from all those
satisfying $E=SES^{-1} $. The $S$ here is expressed as
$S(k_i)=\sum_ik_i|\Psi _i^R\rangle \langle \Psi_i^L|$ by invoking
the spectrum structure of $E$. The distinguishing problem can be
resolved following the same protocol described above.

Below we present an example to illustrate the scheme and construct
the interaction model with presumed symmetry. Consider a family of
MPSs $|\Psi (g)\rangle $ with
\begin{equation}
A^0=\left[
\begin{array}{ll}
1 & 0 \\
0 & g
\end{array}
\right] ,A^{+}=\left[
\begin{array}{ll}
0 & 0 \\
g & 0
\end{array}
\right] ,A^{-}=\left[
\begin{array}{ll}
0 & 1 \\
1 & 0
\end{array}
\right],  \label{example}
\end{equation}
where $g$ is a real parameter. The corresponding $E$ has the largest
eigenvalue $\lambda _1=\frac 12(\Omega +\sqrt{4+\Omega ^2})$ with
$\Omega =1+g^2$. By comparing the spectrum of $E$ and $E^{T_2}$, one
finds that the state is parity absent except cases of $g=0,\pm1$. To
reveal the possible symmetry and hidden string order, it is
instructive to specify the four left eigenvectors of $E$ as
\begin{equation}
|\Psi _{1,2}^L\rangle =\gamma _{1,2}|00\rangle +|11\rangle ,~~|\Psi
_{3,4}^L\rangle =|01\rangle \pm |10\rangle ,  \label{leftei}
\end{equation}
where $\gamma _{1,2}=\frac 12(1-g^2\pm \sqrt{4+\Omega ^2})$. In
terms of bases (\ref{magic}), the transformation
$S(k_i)=\sum_{i=1}^4k_i|L_i\rangle \langle L_i|$ is expressed as
$D^s(k_i)=$(\negthinspace {\tiny $
\begin{array}{ll}
B_1 & 0 \\
0 & B_2
\end{array}
\!\!$}), in which $B_2=$(\negthinspace {\tiny $
\begin{array}{ll}
2k_3 & 0 \\
0 & 2k_4
\end{array}
\!\!$}) and
\begin{equation}
B_1=\frac 12\sum_{i=1}^2\left[
\begin{array}{ll}
k_i(1+\gamma _i)^2 & ik_i(1-\gamma _i^2) \\
ik_i(\gamma _i^2-1) & k_i(1-\gamma _i)^2
\end{array}
\right] .  \label{comrep}
\end{equation}
By substituting it into the condition $D^s(k_i)[D^s(k_i)]^T=I$, one
derives
\begin{equation}
k_1=\frac{\sqrt{-\gamma _1\gamma _2}}{\gamma _1^2-\gamma _1\gamma
_2},~~k_2=\frac{\sqrt{-\gamma _1\gamma _2}}{\gamma _2^2-\gamma
_1\gamma _2},~~k_3=k_4=\frac 12.  \label{ki}
\end{equation}
Thus one can infer that the specified operator $S(k_i)$ falls into
${\rm SL(2,C)}\otimes {\rm SL(2,C)}$ according to the described
group Homomorphism. Simply, the decomposition $S(k_i)=X_1\otimes
X_2$ of this case can be obtained by expressing $S(k_i)$ in the
conventional bases $\{|00\rangle ,|01\rangle ,|10\rangle ,|11\rangle
\}$ as it has a diagonal form $S(k_i)={\rm diag}\{\Omega ^{\frac
12},1,1,\Omega ^{-\frac 12}\}$. The yielded $X_{1,2}$ and $U_P$
satisfying Eq. (\ref{localeq}) turn out to be
\begin{equation}
X_{1,2}={\rm diag}\{\Omega ^{\frac 14},\Omega ^{-\frac 14}\},
~U_P(g)=|0\rangle \langle 0|-ie^{i\frac \pi 2\mathbf{n}\cdot
\mathbf{\sigma }},  \label{xuexpr}
\end{equation}
where the Pauli operator $\mathbf{\sigma }$ is defined in a
two-state space $\{|+\rangle ,|-\rangle \}$ and the reflection is
taken in the Bloch space along $\mathbf{n}=(\cos \theta ,0,-\sin
\theta )$ with $\theta =\arctan \frac 1g$. As a consequence, the
string order parameters ${\rm SO}_{GP}^{\alpha }$ of $|\Psi
(g)\rangle $ are figured out in Fig. 1.

\begin{figure}[tbp]
\begin{center}
\epsfig{figure=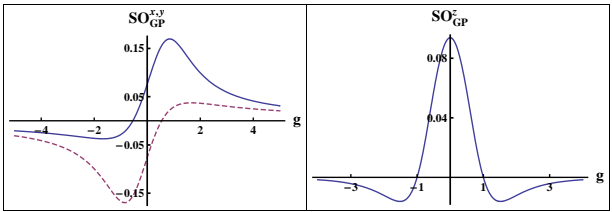,width=0.45\textwidth}
\end{center}
\caption{The transverse string order parameters ${\rm SO}_{GP}^x$
(solid line in the left), ${\rm SO}_{GP}^y$ (dashed line in the
left) and the longitudinal ${\rm SO}_{GP}^z$ (right) in the state
$|\Psi (g)\rangle $ specified by Eq. (\ref{example}). It is shown
that there is ${\rm SO}_{GP}^x(-g)=-{\rm SO}_{GP}^y(g)$.}
\end{figure}

The parent Hamiltonian with $k$-local interactions always exists for
a given MPS as $k>2\log_dD$ \cite{fannes}. It can be constructed by
a sum of positive operators supported in the null space of a
Hermitian matrix $\mathcal{A}\mathcal{A}^{\dagger}$, where
$\mathcal{A}$ is a $d^k\times D^2$ matrix with
$\mathcal{A}^{i_1\cdots i_k}_{\alpha\beta}\equiv(A^{i_1}\cdots
A^{i_k})_{\alpha\beta }$. For the state $|\Psi (g)\rangle $
specified by Eq. (\ref{example}) one gets a Hamiltonian with $k=2$,
i.e., $H=\sum_ih(i,i+1)$. As $g=1$, the detailed form of $h$ is
expressed as
\begin{eqnarray}
h&=&\sum_{\mu =x,y}\left[\{S_i^\mu S_{i+1}^\mu
,S_i^zS_{i+1}^z\}_{+}+S_i^\mu S_{i+1}^\mu (S_i^z+S_{i+1}^z)\right]
\nonumber \\
&&-\frac 12S_i^zS_{i+1}^z-\frac 12(A_iB_{i+1}+B_iA_{i+1})+\frac 12K,
\label{hamil}
\end{eqnarray}
where $A=(S^x)^2-(S^y)^2$, $B=(S^z)^2-S^z$ and
$K=(S_i^zS_{i+1}^z)^2+(S_i^{+}S_{i+1}^{-})^2+(S_i^{-}S_{i+1}^{+})^2$.
The state $|\Psi (1)\rangle $ turns to be a unique ground state of
the model by construction \cite{fannes,wolf} and it has dual
symmetries under the parity and local spin reflection $U_P(1)$ [cf.
Eq. (\ref{xuexpr})]. Consequently, the system at this point has
non-vanishing string order via both passages of the parity and the
local unitary, e.g., they are obtained as ${\rm SO}_P^x=\frac 38$,
and ${\rm SO}_D^x=\frac {\sqrt{2}}8$, respectively.

Summing up, the symmetry related to spatial reflection has been
investigated for quantum spin chains and a set of renewed string
order was proposed. The quantity is shown applicable to characterize
a variety of physical systems owing to the particular role of
spatial reflection in spin lattice systems. Furthermore, revelation
of the generalized parity symmetry is important in many aspects.
Besides that it can be exploited to reduce computational costs in
numerical algorithm, we build an explicit scheme to determine it for
MPSs so that one can construct ansatz models with presumed
properties. In addition, the string correlation was shown related to
the concept of localizable entanglement \cite{locentan}, a quantity
reflecting the quality of a quantum channel. It would be of interest
to explore further the role of the string order variants in the
context of quantum information theory.

The author thanks Peng Li for helpful discussions. This work was supported
by the NSFC under grant Nos. 10604043 and 10874254.

\end{document}